\documentclass[conference]{IEEEtran}

\usepackage[T1]{fontenc}
\usepackage[utf8]{inputenc}
\usepackage{cite}
\usepackage{amsmath,amssymb}
\usepackage{graphicx}
\usepackage{booktabs}
\usepackage{url}
\usepackage{hyperref}
\usepackage{microtype}

\graphicspath{{images/}}

\begin{document}

\title{A Longitudinal Study of Recently Observed Malicious Domains:\\
Characteristics, Infrastructure, and Abuse Patterns}

\author{%
  \IEEEauthorblockN{Fathima Mashood, Mohamed Nabeel}
  \IEEEauthorblockA{Southern New Hampshire University, National University\\
  fathima.mashood@snhu.edu, mmohamednabeel@nu.edu}
}

\maketitle

\begin{abstract}
We present a longitudinal study of approximately 1.52 million malicious
domains observed on VirusTotal (VT) between January and May 2026.
Domains were selected on the basis of detection by at least five independent
VT scanning engines and a first-seen date within the study window.
We group the dataset into \emph{compromised} domains and \emph{attacker-created}
domains, which account for approximately 89.3\% of the dataset.
Combining WHOIS registration records and passive DNS (PDNS) data with the VT
dataset, we characterise attacker behaviour across eight dimensions:
temporal distribution, compromised-vs.-attack classification, domain age at
first detection, registrar and TLD preferences, DNS query volume as a damage
proxy, hosting infrastructure concentration (IP and ASN level), bulk
registration patterns, and brand impersonation.
Key findings include: the majority of attacker-created domains are short-lived
registrations used within weeks of creation; a small number of registrars and
TLDs account for most abuse; Cloudflare infrastructure is heavily exploited
for domain fronting; bulk registration events involving thousands of domains from a
single registrar on a single day are widespread; and several global brands,
particularly WhatsApp and Google, are heavily impersonated. We share the annotated dataset in the GitHub repo \url{https://github.com/mufimash/malicious_domains} for further research.
\end{abstract}

\section{Introduction}
\label{sec:intro}

Malicious domains are the primary delivery mechanism for phishing campaigns,
malware distribution, command-and-control (C2) communications, and online
fraud.
Despite continuous blocklisting efforts, attackers register new domains at a
high rate, exploiting low-cost registration services, privacy-protection
features, and a wide choice of generic TLDs (gTLDs) to evade detection
\cite{felegyhazi2010,hao2013}.
Understanding the characteristics of these domains on a scale is essential for
improving detection, prioritizing takedown efforts, and informing policy.

Prior work has studied malicious domain behavior from several angles,
including DNS-based reputation systems \cite{antonakakis2012}, network-level
traffic patterns \cite{lever2016}, and brand-squatting typologies
\cite{kintis2017}.
However, the rapid evolution of the threat landscape means that studies older
than a few years may not reflect current attacker tactics. Additionally, these studies lack longitudinal understanding of malicious domains in general.

This paper addresses this gap with a contemporaneous dataset of approximately
1.52 million malicious domains first observed by VirusTotal between January
and May 2026. While VT is not comprehensive, this platform is used widely by users all of the world covering a wide range of malicious domains.
We annotate malicious domains with WHOIS registration data, PDNS resolution histories,
and the Tranco top-1M popularity list \cite{tranco2019} to answer the
following research questions. In order to better characterize malicious domains, based on prior research \cite{compromised2021}, we use simplified heuristics to first categorize malicious domains as compromised or attacker-create. A compromised domain is a legitimate domain that attackers have exploited to launch an attack whereas attacker-created domains are malicious registered by attackers. This distinction is important in order not to skew the statistical insights we obtain answering the following research questions.

\begin{itemize}
  \item \textbf{RQ1:} How are newly detected malicious domains distributed
    over time, and what fraction are attacker-created versus compromised?
  \item \textbf{RQ2:} How quickly after registration do attacker-created
    domains appear in VT?
  \item \textbf{RQ3:} Which registrars and TLDs are most heavily abused?
  \item \textbf{RQ4:} What does PDNS query volume reveal about the scale of
    harm caused?
  \item \textbf{RQ5:} How concentrated is the hosting infrastructure, and do
    attackers share IP and ASN resources?
  \item \textbf{RQ6:} Is there evidence of coordinated bulk registration?
  \item \textbf{RQ7:} Which popular brands are most frequently impersonated?
\end{itemize}

The rest of the paper is organized as follows. Section~\ref{sec:related} reviews related work.
Section~\ref{sec:data} describes the dataset and annotation methodology.
Section~\ref{sec:findings} presents our findings.
Section~\ref{sec:conclusion} concludes.

\section{Related Work}
\label{sec:related}

Antonakakis~et~al.~\cite{antonakakis2012} proposed Notos, a dynamic DNS
reputation system that scores domains by aggregating features derived from
passive DNS data, demonstrating that passive DNS provides rich behavioural
signals that distinguish malicious domains from benign ones.
We build on this insight by using PDNS query counts as a proxy for harm.

Hao~et~al.~\cite{hao2013} studied the initial DNS registration and
resolution behaviour of malicious domains, finding that attackers exploit
newly registered domains rapidly.
Their methodology motivates our focus on domains with a 2026 first-seen date
in VT, capturing domains at the earliest stage of their malicious lifecycle.

Lever~et~al.~\cite{lever2016} conducted a five-year longitudinal analysis of
malware network communications, characterising how attackers reuse and share
hosting infrastructure.
We extend this perspective to 2026 and confirm at the IP and ASN level that
infrastructure concentration remains a defining attacker behaviour.

Kintis~et~al.~\cite{kintis2017} introduced the concept of combosquatting --
embedding a popular brand name into a domain name alongside extra terms -- and
showed it is widespread and persistent.
Our brand impersonation analysis uses substring matching against Tranco top-10K
brands, filtered against a common-word corpus to reduce noise.

Marchal~et~al.~\cite{marchal2016} proposed PhishStorm, a streaming analytics
approach to detecting phishing domains based on lexical and structural
features, highlighting the value of large-scale, real-time domain analysis
aligned with our VirusTotal-based collection methodology.

De Silva~et~al.~\cite{compromised2021} proposed a content agnostic method to classify malicious domains as either compromised or attacker-created. While they used VT data, their study was limited to two relatively small datasets extracted from two different time periods of one week. 

To the best of our knowledge, this is the first large-scale study that characterizes VT observed malicious domains over a 5 month period in 2026.
\section{Data Collection and Annotation}
\label{sec:data}

\subsection{Primary Dataset}

The primary dataset was obtained from VirusTotal (VT) \cite{virustotal2024}.
We collected all registered domain (i.e. second level domain names) records for which (a) at least five independent
scanning engines flagged the domain as malicious, and (b) the first-seen date
recorded by VT falls within January and May 2026. The threshold 5 was selected based on the common practice followed by prior research and the security industry.
The resulting dataset contains 1{,}520{,}050 unique domain names.

\subsection{Annotation Sources}

\textbf{Tranco top-1M.}
The Tranco list \cite{tranco2019} provides a research-oriented,
manipulation-resistant ranking of popular domains.
We use it for compromised-domain classification and brand impersonation
analysis. Since Tranco popularity does not always translate to benignity, our study may exclude some attacker-created popular domains. However, the percentage of such domains is below 1\% and hence the impact on the study is minimal.

\textbf{WHOIS records.}
WHOIS data provides registration information such as registrar, creation date, expiry date, and nameserver
information.
Approximately 10.1\% of records lack a parseable \texttt{created\_date} and
are handled with Tranco-only classification. This is a known limitation of WHOIS records and 10\% missing WHOIS records is in par with previous studies.

\textbf{Passive DNS (PDNS).}
PDNS records capture observed DNS resolutions, providing IP addresses, ASNs,
query counts, and temporal coverage. PDNS records are extracted from FarSight (acquired by Domain Tools) PDNS database. Approximately 11.9\% of records contained an empty IP set, which is expected as FarSight does not have visibility to certain IP ranges.

\subsection{Dataset Statistics}

Table~\ref{tab:stats} summarises coverage after annotation.

\begin{table}[t]
  \centering
  \caption{Dataset Statistics}
  \label{tab:stats}
  \begin{tabular}{lr}
    \toprule
    \textbf{Metric} & \textbf{Count} \\
    \midrule
    Total VT malicious domains & 1{,}520{,}050 \\
    Compromised domains & 162{,}129 (10.7\%) \\
    \quad -- by Tranco rank $\leq$ 500K only & 2{,}838 \\
    \quad -- by age $\geq$ 3 years only & 157{,}707 \\
    \quad -- by both conditions & 1{,}584 \\
    Attack domains & 1{,}357{,}921 (89.3\%) \\
    Domains with WHOIS created\_date & $\approx$88.4\% \\
    Domains with non-empty PDNS IP set & $\approx$88.1\% \\
    \bottomrule
  \end{tabular}
\end{table}

\subsection{Temporal Distribution}

\begin{figure}[t]
  \centering
  \includegraphics[width=\columnwidth]{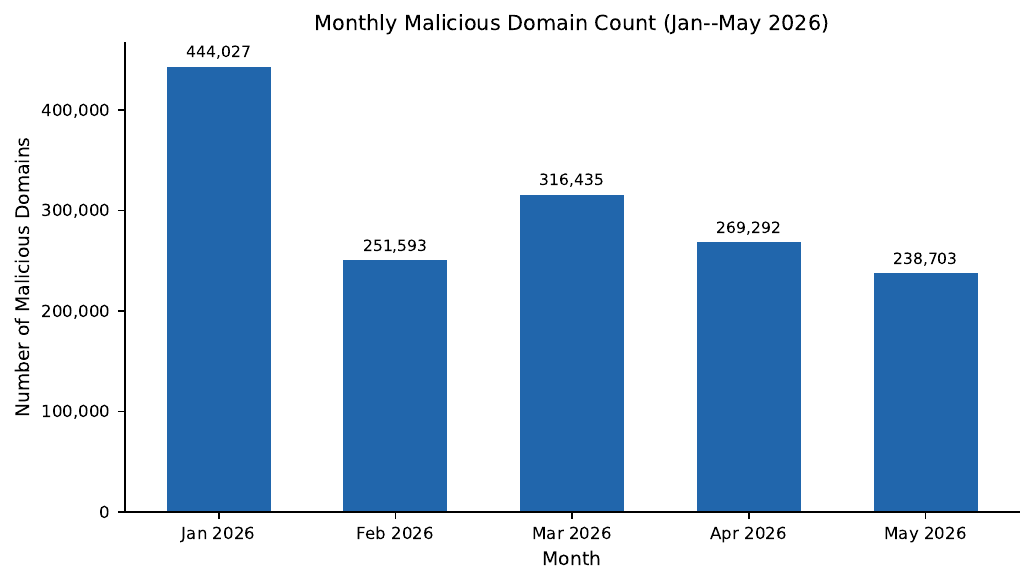}
  \caption{Monthly count of malicious domains first detected on VirusTotal
    (January to May 2026).}
  \label{fig:monthly}
\end{figure}

Figure~\ref{fig:monthly} shows the number of malicious domains first seen on
VT by calendar month.
January 2026 recorded the highest count (444{,}027), with subsequent months
showing moderate variation between 238{,}703 and 316{,}435 domains.
The consistently high monthly volumes of newly observed malicious domains confirm that malicious domain creation is a sustained, large-scale activity. This finding is consistent with prior research that attackers use disposable domains to launch attacks.

\section{Findings and Implications}
\label{sec:findings}

\subsection{Compromised vs.\ Attacker-Created Domains}
\label{sec:classify}

We classify a VT-flagged domain as \emph{compromised} if either (a) it
appears in the Tranco top-1M list with a rank of 500{,}000 or below, or (b)
its registration age at first VT detection is three years or more
(i.e., the WHOIS creation date precedes the first-seen date by at least
1{,}095 days).
The first condition identifies hijacked popular websites; the second captures
legitimate older domains that may have been compromised long
after initial registration. As mentioned earlier, this classification may miss some popular malicious domains or strategically aged malicious domains from the attacker-created domains. However, it should not affect the findings and implications of the study as the results are discussed based on a very conservative lower bound.
All remaining domains are classified as \emph{attacker-created}.

Of the 1{,}520{,}050 total domains, 162{,}129 (10.7\%) are classified as
compromised: 2{,}838 by Tranco rank alone, 157{,}707 by age alone, and 1{,}584
by both criteria.
The remaining 1{,}357{,}921 domains (89.3\%) are attacker-created.
The predominance of the age condition in the compromised class reveals that
many long-established domains have been repurposed or abused, whereas freshly
created infrastructure dominates the attack domain class.

\subsection{Attack Domain Age}
\label{sec:age}

Domain age is defined as the number of days between the WHOIS
\texttt{created\_date} and the VT \texttt{first\_seen} timestamp.
Domains with missing creation dates or negative ages are excluded, leaving
1{,}158{,}280 attack domains for this analysis.

\begin{figure}[t]
  \centering
  \includegraphics[width=\columnwidth]{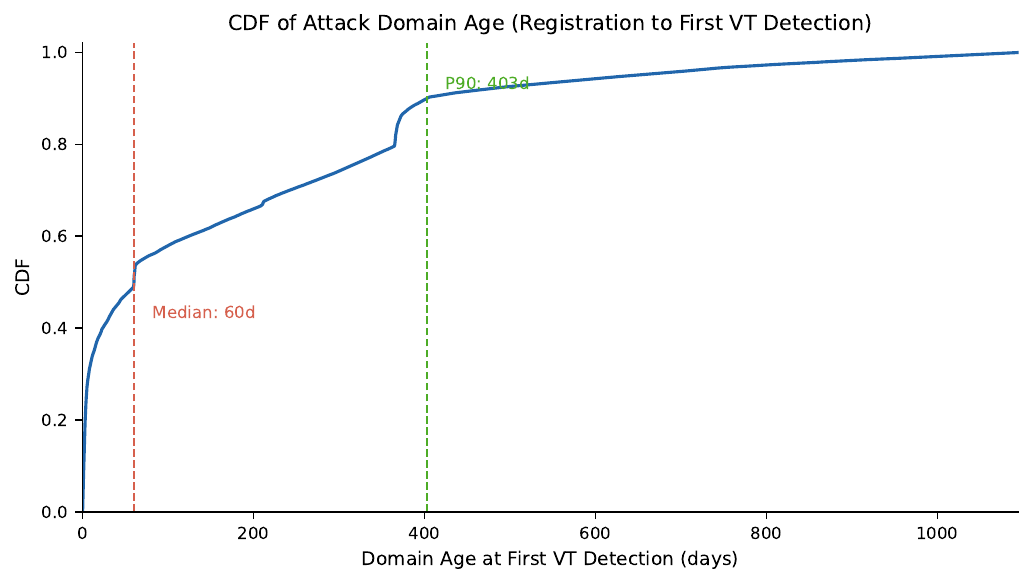}
  \caption{CDF of attack domain age (days from WHOIS creation to first VT
    detection), x-axis capped at 3 years (1{,}095 days).
    Median = 60 days; P90 = 403 days.}
  \label{fig:age_cdf}
\end{figure}

Figure~\ref{fig:age_cdf} shows the CDF capped at three years (1{,}095 days).
The median age is 60 days, and the 90th percentile is 403 days (approximately
13 months).
Approximately 7.4\% of attack domains were detected within one day of
registration, and 29.9\% within one week.
This rapid deployment window leaves defenders very little time for proactive
identification before the domains are activated for malicious use.

\subsection{Domain Registrars}
\label{sec:registrars}

\begin{figure}[t]
  \centering
  \includegraphics[width=\columnwidth]{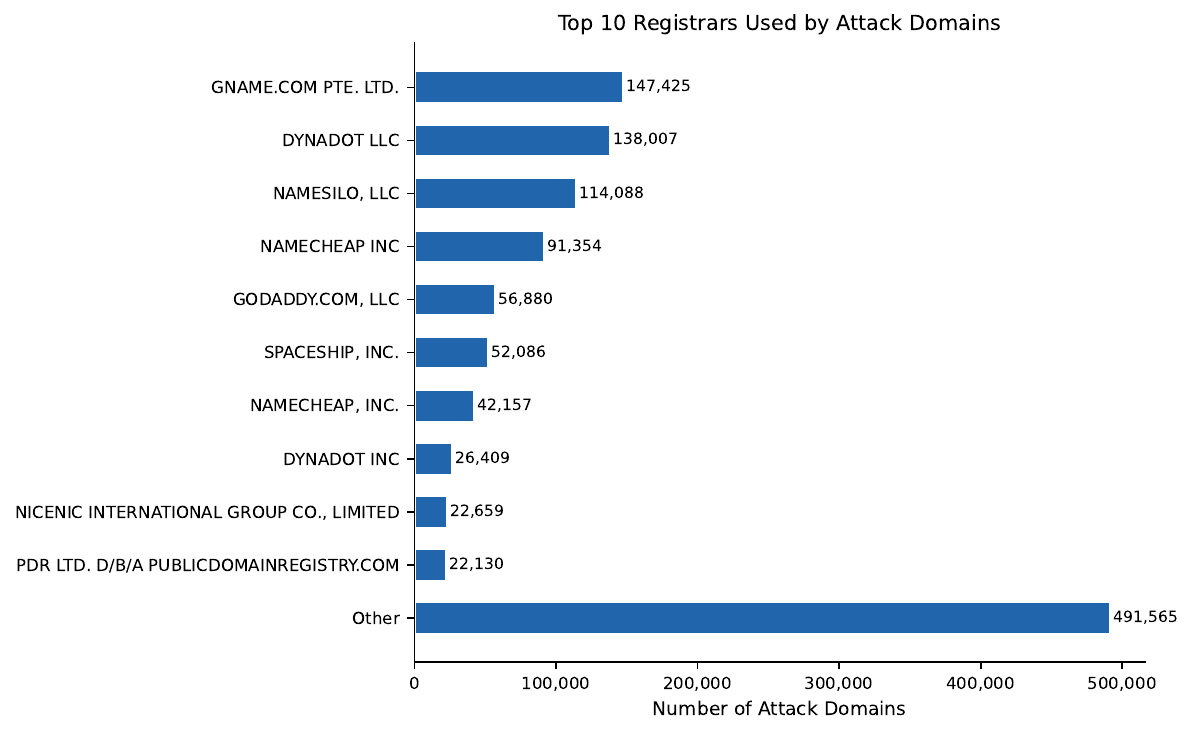}
  \caption{Top 10 domain registrars used by attack domains.
    An ``Other'' bar aggregates the remaining 4{,}464 registrars.}
  \label{fig:registrars}
\end{figure}

Figure~\ref{fig:registrars} presents the top 10 registrars by number of
attack domains.
GNAME.COM (147{,}425), Dynadot LLC (138{,}007), Namesilo LLC (114{,}088), and
Namecheap Inc (91{,}354) together account for more than 36\% of all attack
domains.
The top 10 registrars together handle 59.2\% of attack domains with known
registrar data, while the long tail spans 4{,}464 additional registrars
comprising the remaining 40.8\%.

This concentration has practical implications: engaging the top 10 registrars
with enhanced anti-abuse policies and accelerated takedown procedures could
substantially reduce the volume of attacker-created domains.

\subsection{Top-Level Domain Abuse}
\label{sec:tlds}

\begin{figure}[t]
  \centering
  \includegraphics[width=\columnwidth]{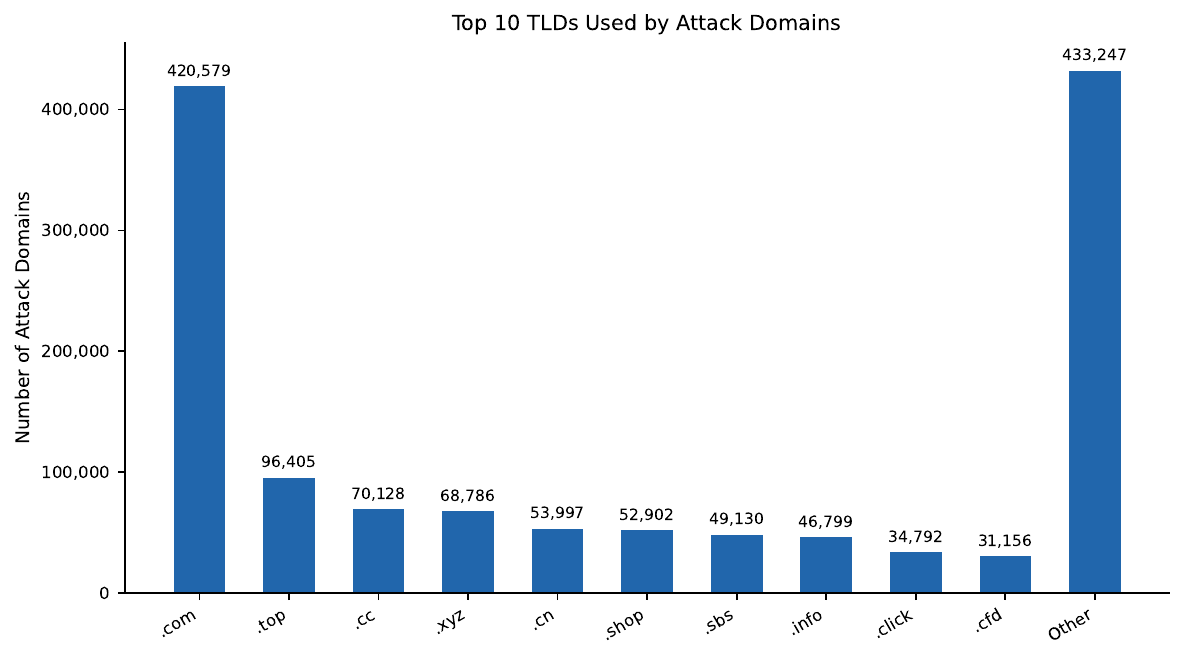}
  \caption{Top 10 TLDs abused by attack domains.}
  \label{fig:tlds}
\end{figure}

Figure~\ref{fig:tlds} shows the top 10 TLDs observed among attack domains.
\texttt{.com} dominates with 420{,}579 domains (31.0\%), followed by
\texttt{.top} (7.1\%), \texttt{.cc} (5.2\%), and \texttt{.xyz} (5.1\%).
The presence of several low-cost gTLDs and country-code TLDs (ccTLDs) in the
top 10 confirms that attackers exploit any registry offering favourable
economics.
The top 10 TLDs account for 68.1\% of all attack domains, while 741 other
TLDs make up the long tail.

\subsection{DNS Query Volume as a Damage Proxy}
\label{sec:pdns}

\begin{figure}[t]
  \centering
  \includegraphics[width=\columnwidth]{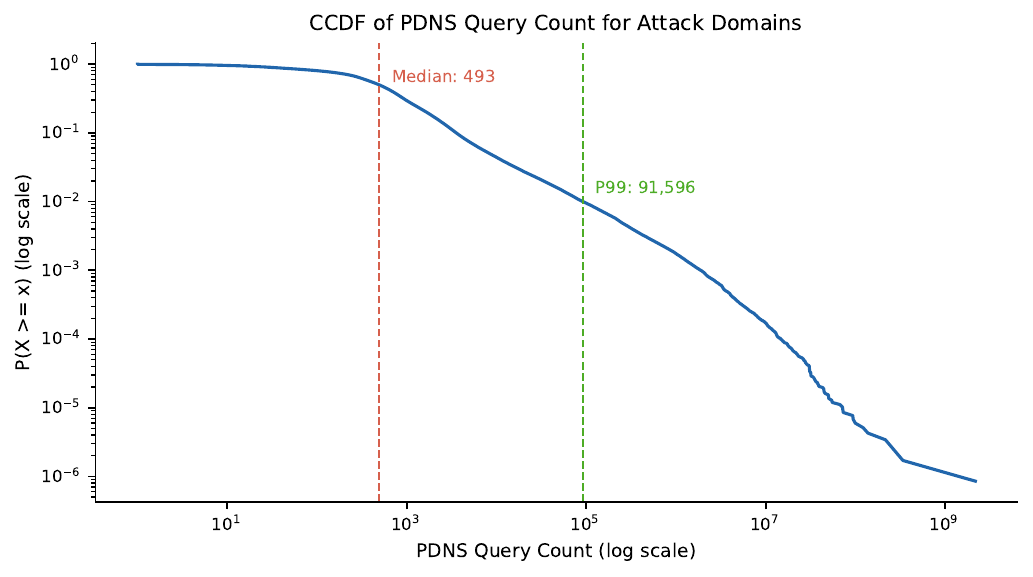}
  \caption{CCDF of PDNS query count for attack domains (log-log scale).}
  \label{fig:ccdf}
\end{figure}

Figure~\ref{fig:ccdf} shows the CCDF of \texttt{pdns\_count} on log-log axes
for 1{,}172{,}386 attack domains with non-zero PDNS records.
The distribution is highly skewed: the median count is 493 queries, while the
99th percentile reaches 91{,}597 and the maximum exceeds 2.1 billion.
The power-law-like shape indicates that a small subset of domains accumulate
the majority of DNS queries and thus pose the greatest harm.
Rapid sinkholing of this high-volume tail would significantly reduce end-user
exposure.

\subsection{Hosting Infrastructure and Cohosting}
\label{sec:hosting}

\subsubsection{IP-Level Hosting}

\begin{figure}[t]
  \centering
  \includegraphics[width=\columnwidth]{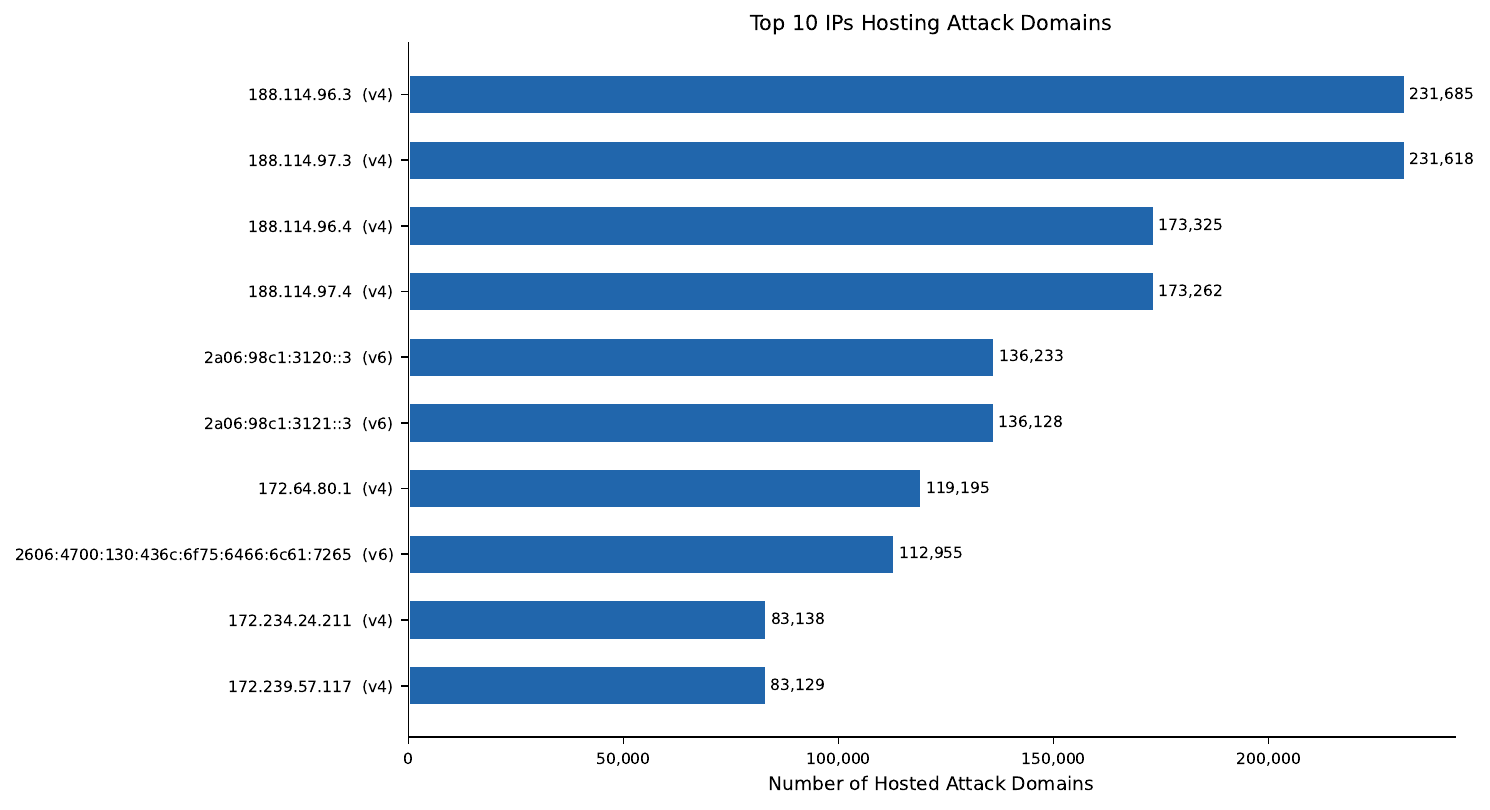}
  \caption{Top 10 IP addresses hosting the most attack domains.
    Bars are labelled with IP version (v4/v6).}
  \label{fig:hosting_bar}
\end{figure}

Figure~\ref{fig:hosting_bar} presents the top 10 IP addresses hosting the
largest numbers of distinct attack domains.
Eight of the top 10 belong to Cloudflare (four IPv4, four IPv6 addresses in
the \texttt{188.114.0.0/16}, \texttt{2a06:98c1::/32}, and
\texttt{172.64.0.0/16} ranges).
The two leading IPv4 addresses each host over 230{,}000 distinct attack
domains, reflecting widespread use of Cloudflare's reverse-proxy and CDN
services to mask the true origin of malicious content.

\begin{figure}[t]
  \centering
  \includegraphics[width=\columnwidth]{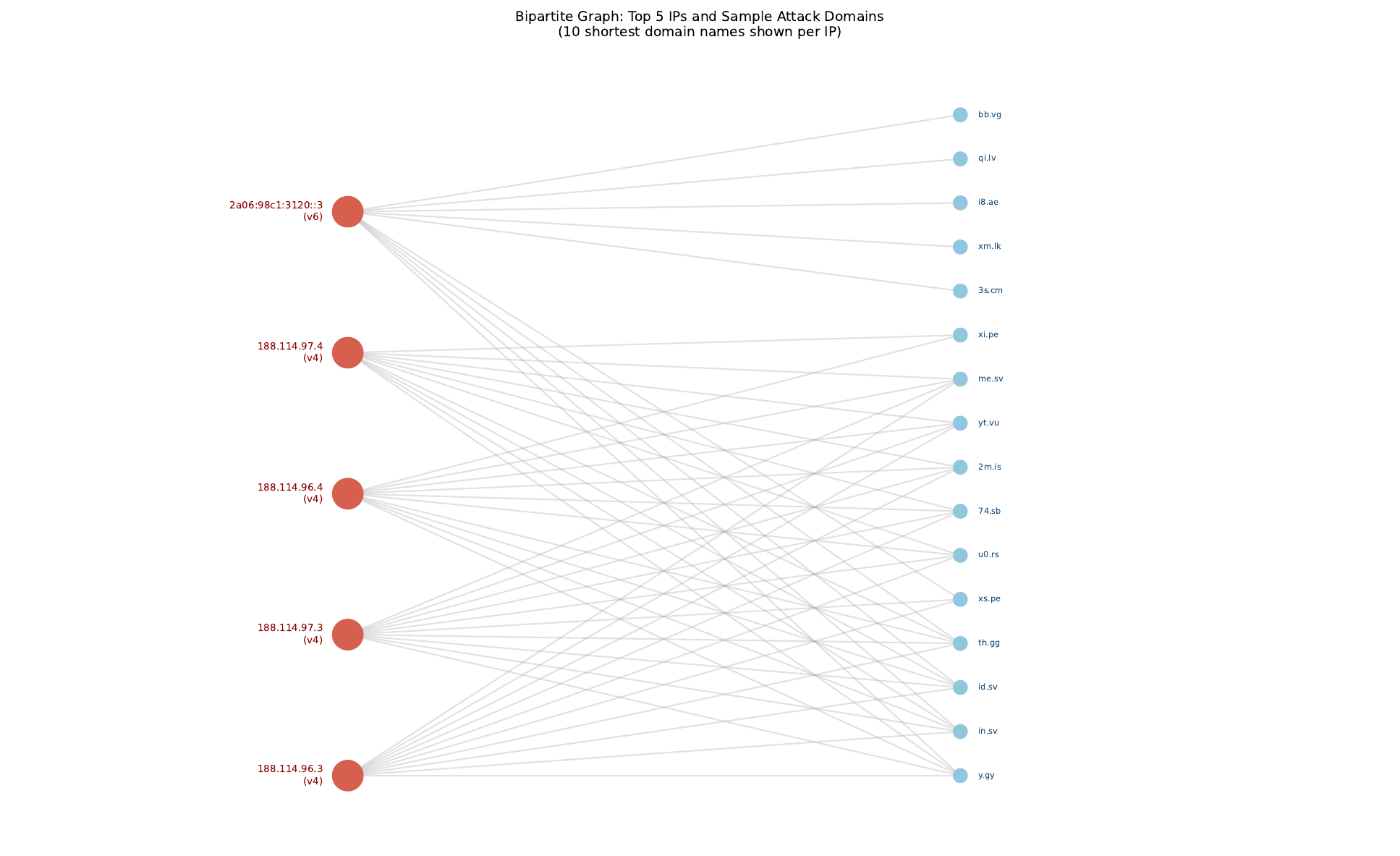}
  \caption{Bipartite network of the top 5 hosting IPs (red) and sample attack
    domains (blue), with domain name labels visible.}
  \label{fig:hosting_net}
\end{figure}

Figure~\ref{fig:hosting_net} visualizes the bipartite domain-IP relationship
for the five most heavily used IPs, with both IP and domain name labels shown.
The hub-and-spoke structure confirms that a small number of IP addresses serve
as shared malicious hosting platforms \cite{noroozian2016}.

\subsubsection{ASN-Level Hosting}

\begin{figure}[t]
  \centering
  \includegraphics[width=\columnwidth]{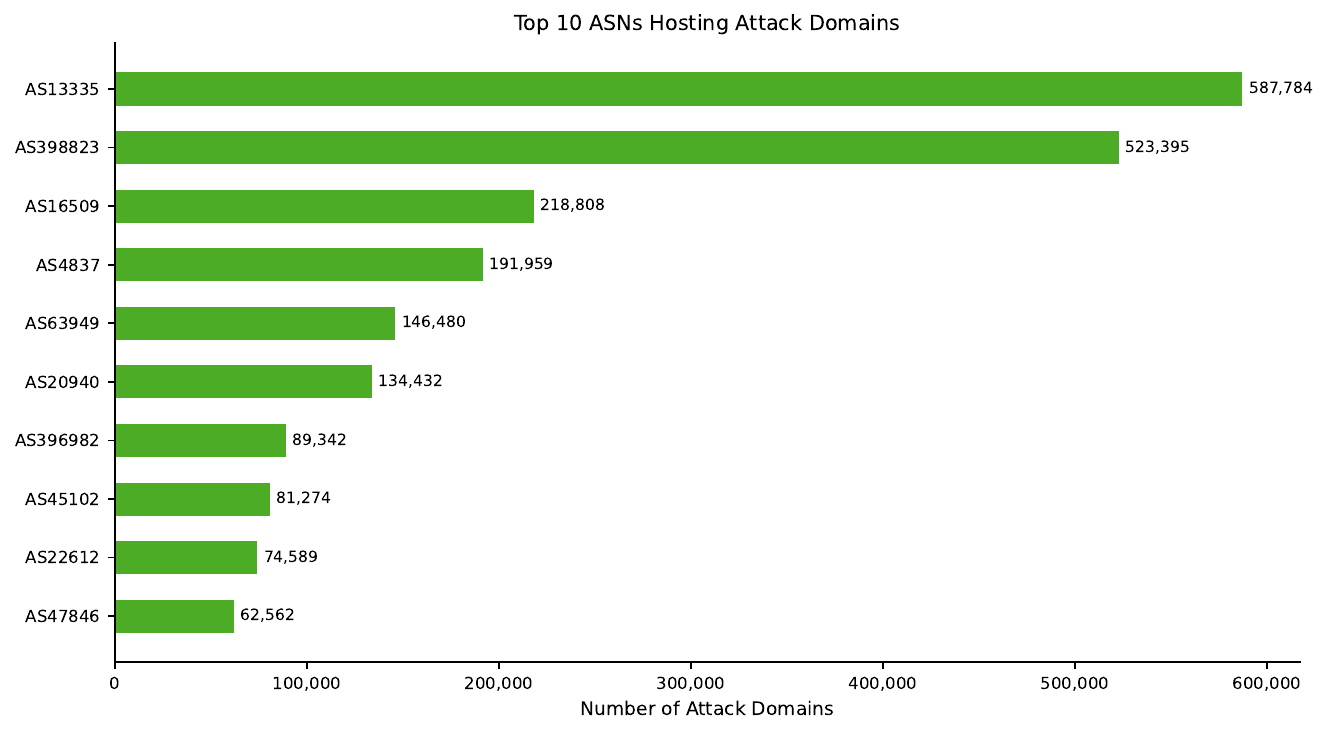}
  \caption{Top 10 ASNs hosting attack domains by number of distinct domains.}
  \label{fig:asns}
\end{figure}

Figure~\ref{fig:asns} shows the top 10 autonomous systems (ASNs) by number
of distinct attack domains.
AS13335 (Cloudflare) hosts 587{,}784 attack domains, followed by AS398823
(523{,}395) and AS16509 (Amazon AWS, 218{,}808).
The dominance of major cloud and CDN providers in this ranking reflects
attackers' preference for reputable infrastructure to reduce the likelihood
of network-level blocking.
Providers such as Cloudflare and AWS offer abuse reporting mechanisms, but the
scale of misuse suggests that automated takedown pipelines coordinated between
threat intelligence teams and these networks are needed.

\subsection{Bulk Registration Patterns}
\label{sec:bulk}

To identify coordinated registration activity, we group attack domains by
(registrar, WHOIS creation date).
A \emph{batch} is defined as a group containing at least five domains sharing
these two attributes.
Among 1{,}357{,}921 attack domains with usable WHOIS records, 1{,}057{,}926
(77.9\%) belong to a batch, spanning 22{,}977 distinct batches across 106{,}137
unique (registrar, date) combinations.

Table~\ref{tab:bulk} shows a selection of the largest batches.
The top batch alone contains 2{,}168 domains registered with Namesilo LLC on
2026-02-06.
Domain names within batches commonly follow short alphanumeric patterns
(e.g., \texttt{10jfr.top}, \texttt{10jgr.top}) indicative of programmatic
generation.
Batches occurring on the same day under the same registrar suggest automated
mass registration, likely driven by scripts that generate and register domain
names in bulk to build large attack fleets quickly.

\begin{table}[t]
  \centering
  \caption{Sample of Largest Bulk Registration Batches}
  \label{tab:bulk}
  \begin{tabular}{clcr}
    \toprule
    \textbf{Rank} & \textbf{Registrar} & \textbf{Date} & \textbf{Size} \\
    \midrule
    1  & Namesilo, LLC      & 2026-02-06 & 2{,}168 \\
    2  & Namecheap Inc      & 2025-01-06 & 2{,}017 \\
    3  & Namesilo, LLC      & 2026-03-09 & 1{,}984 \\
    4  & Namecheap Inc      & 2025-02-05 & 1{,}850 \\
    5  & Namecheap Inc      & 2025-01-07 & 1{,}536 \\
    6  & Spaceship, Inc.    & 2026-01-13 & 1{,}432 \\
    7  & Dynadot LLC        & 2025-07-11 & 1{,}402 \\
    8  & Spaceship, Inc.    & 2026-01-10 & 1{,}392 \\
    9  & Dynadot LLC        & 2026-02-17 & 1{,}383 \\
    10 & GNAME.COM Pte Ltd  & 2025-09-03 & 1{,}362 \\
    \bottomrule
  \end{tabular}
\end{table}

\subsection{Brand Impersonation}
\label{sec:brands}

To quantify brand abuse, we extract brand tokens from the Tranco top-10{,}000
domains.
Brand tokens are the second-to-last domain component (e.g., \texttt{paypal}
from \texttt{paypal.com}), restricted to at least four characters and filtered
against the system English word corpus  to
remove common words such as ``live,'' ``shop,'' and ``cloud.''
After filtering, 7{,}154 brand tokens are retained for matching.

We test whether each brand token appears as a substring within an attack
domain's registered name (excluding the TLD), capturing both combosquatting
\cite{kintis2017} and typosquatting variants.

\begin{figure}[t]
  \centering
  \includegraphics[width=\columnwidth]{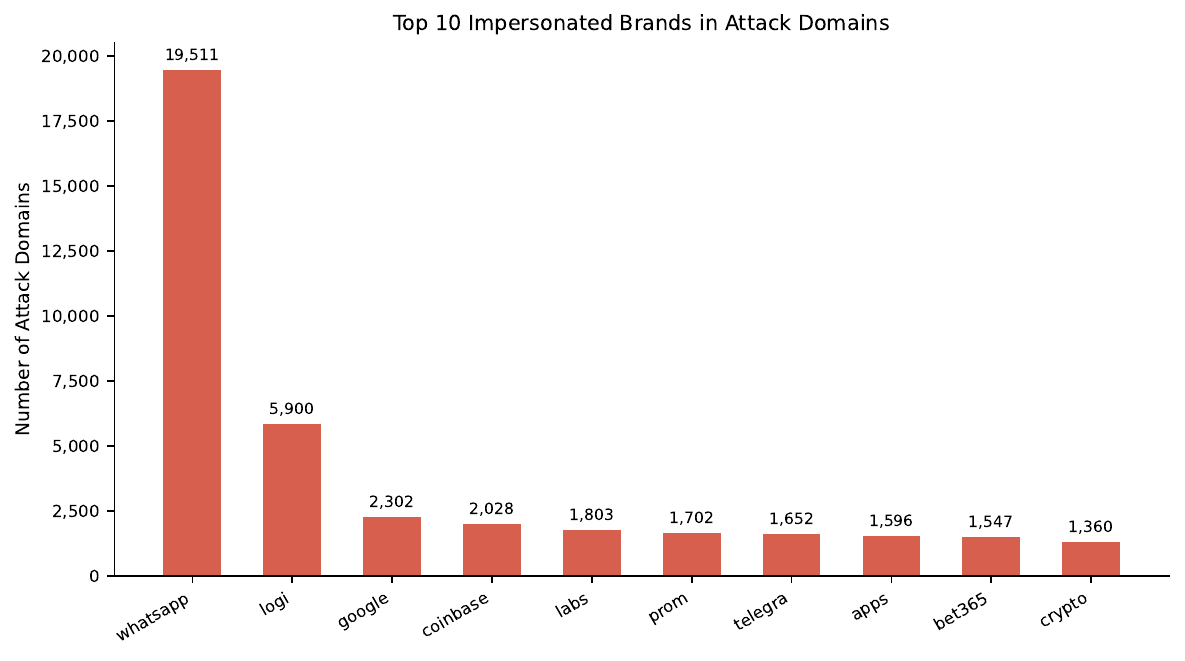}
  \caption{Top 10 brands impersonated in attack domains (brand substring
    match, common English words excluded).}
  \label{fig:brands}
\end{figure}

Figure~\ref{fig:brands} shows the top 10 impersonated brands.
WhatsApp is the most abused brand with 19{,}511 attack domains containing the
token ``whatsapp,'' followed by Logitech (\texttt{logi}: 5{,}900) and Google
(2{,}302).
The presence of Coinbase (2{,}028) and Bet365 (1{,}547) highlights that
cryptocurrency exchanges and gambling platforms are also heavily targeted,
likely for credential-harvesting and phishing.
In total, 114{,}034 attack domains (8.4\%) contain a brand token, and 2{,}569
distinct brands are impersonated at least once.

\section{Conclusion}
\label{sec:conclusion}

This paper presented a large-scale analysis of approximately 1.52 million
malicious domains first observed on VirusTotal between January and May 2026.
Our analysis yielded eight principal findings:

\begin{itemize}
  \item \textbf{Attacker-created domains dominate (89.3\%).}
    The majority of flagged domains are purpose-registered, with the remaining
    10.7\% classified as compromised primarily because they are long-established
    domains repurposed for malicious use.
  \item \textbf{Rapid deployment after registration.}
    Median attack domain age at first VT detection is only 60 days;
    7.4\% are detected within one day of registration.
  \item \textbf{Registrar and TLD concentration.}
    Four registrars account for over 36\% of attack domains; the top
    10 TLDs handle 68\% of abuse.
  \item \textbf{Skewed DNS traffic distribution.}
    A small tail of high-volume domains with millions of PDNS queries
    accounts for disproportionate harm and should be prioritised for
    sinkholing.
  \item \textbf{Cloudflare infrastructure predominates.}
    Eight of the top 10 hosting IPs and the leading ASN (AS13335) belong to
    Cloudflare, reflecting attackers' reliance on reputable CDN infrastructure.
  \item \textbf{Automated bulk registration is pervasive.}
    Over 77\% of attack domains with WHOIS data belong to bulk batches; the
    largest single batch comprises 2{,}168 domains registered in one day.
  \item \textbf{Pervasive brand impersonation.}
    8.4\% of attack domains impersonate a known brand; WhatsApp, Google, and
    Coinbase are the top targets.
\end{itemize}

\noindent
These findings point to concrete intervention opportunities: registrar-level
rate limits and anti-abuse reviews for same-day bulk registrations; DNS
provider collaboration to flag domains with known malicious IP footprints;
automated brand-monitoring pipelines for high-value targets; and coordinated
sinkholing for the highest-query-volume attack domains.

\bibliographystyle{IEEEtran}
\bibliography{refs}

\end{document}